\ifdefined\PRD
\documentclass[usenatbib, twocolumn, nofootinbib, reprint,]{revtex4-1}
\else
\documentclass[usenatbib, twocolumn, nofootinbib]{aastex62}
\fi
\usepackage{newtxtext}
\usepackage[normalem]{ulem}
\usepackage{multirow}
\usepackage{ae,aecompl,amsmath,amsfonts,bm}
\usepackage{hyperref}
\usepackage{natbib	}
\usepackage[T1]{fontenc}
\usepackage{amssymb}
\usepackage{booktabs}
\usepackage{color}
\usepackage{dcolumn}
\usepackage{graphicx}
\usepackage{epstopdf}
\usepackage{MnSymbol}
\usepackage{mathtools}

\usepackage{graphicx}	
\usepackage{amsmath}	
\usepackage{amssymb}

\usepackage{dcolumn}
\newcolumntype{C}[1]{>{\centering\let\newline\\\arraybackslash\hspace{0pt}}m{#1}}

\ifdefined\PRD

\newcommand{\aap}{A\&A}

\newcommand{\apjs}{\apj S}

\newcommand{\arcmin}{\ensuremath{^{\prime}}}

\fi

\newcommand{\msolar}{\ensuremath{\mbox{M}_{\odot}}}

\newcommand{\snr}{\ensuremath{S/N}}
\newcommand{\bL}{\bm{\ell}}
\newcommand{\bnhat}{\bm{\hat{\mbox{n}}}}

\newcommand{\sqdeg}{\ensuremath{{\rm deg}^2}}
\newcommand{\cmmnt}[1]{}

\newcommand{\mvir}{M_{200m}}
\newcommand{\msol}{M_{\odot}}
\newcommand{\munits}{\times 10^{14}\ \msol}
\newcommand{\ukarcmin}{\ensuremath{\mu}{\rm K-arcmin}}

\begin{document}

\title{Suppressing the thermal SZ-induced variance in CMB-cluster lensing estimators}
\author{Sanjaykumar Patil}
\affiliation{School of Physics, University of Melbourne, Parkville, VIC 3010, Australia}
\author[0000-0003-1405-378X]{Srinivasan Raghunathan}
\affiliation{Department of Physics and Astronomy, University of California, Los Angeles, CA, USA 90095}\affiliation{School of Physics, University of Melbourne, Parkville, VIC 3010, Australia}
\author[0000-0003-2226-9169]{Christian L. Reichardt}
\affiliation{School of Physics, University of Melbourne, Parkville, VIC 3010, Australia}
\correspondingauthor{Sanjaykumar Patil}
\email{s.patil2@student.unimelb.edu.au}
\keywords{cosmic background radiation -- galaxies: clusters: general -- gravitational lensing: weak}


\begin{abstract}
Accurate galaxy cluster mass measurements from the gravitational lensing of the cosmic microwave background temperature maps depend on mitigating potential biases from the cluster's own thermal Sunyaev-Zel'dovich (SZ) effect signal. 
Quadratic lensing estimators use a pair of maps to extract the lensing signal: a large scale gradient map and a small scale lensing map. 
The SZ bias can be eliminated by using an SZ-free map in the pair, with the gradient map being favored for signal-to-noise reasons. 
However, while this approach eliminates the bias, the SZ power in small scale lensing map adds extra variance that can become significant for high mass clusters and low noise surveys.  
In this work, we propose projecting out an SZ template to reduce the SZ variance.
Any residual SZ signal after template fitting  is uncorrelated with the SZ-free gradient map, and thus does not bias the mass measurements.   
For massive clusters above $4\times 10^{14}$\,\msolar{} observed by the upcoming CMB-S4 and Simons Observatory experiments, we find that the template fitting approach would increase the cluster lensing signal-to-noise by a factor of 1.4. 

\end{abstract}

\ifdefined\PRD
\maketitle
\fi

\section{\noindent Introduction}

 Clusters of galaxies are the largest gravitationally bound objects in the universe.
Their abundance as a function of mass and redshift provides crucial information about the cosmic acceleration and structure formation histories \citep{allen11,mantz08,rozo10,mantz15, dehaan16, salvati17,bocquet18,zubeldia19,vikhlinin09b,hasselfield13}, however, they are currently limited by uncertainties in mass estimation.
While there are a number of survey mass-observable relations, all of these depend on complex baryonic physics which is not well understood.
On the other hand, gravitational lensing is sensitive to the total matter content present inside the cluster irrespective of its dynamical state \citep{linden14,appelgate14}.

Measuring the mass of galaxy clusters using \cmmnt{through} their gravitational lensing imprint upon the cosmic microwave background (CMB) is emerging as a key tool for cluster cosmology, particularly for high-redshift galaxy clusters. 
This technique, called CMB-cluster lensing \citep{seljak00b, holder04, dodelson04,maturi05,lewis06a,hu07,yoo08,melin15,horowitz19}, has advanced rapidly over the last few years, from the initial detections in 2015 \citep{madhavacheril15, baxter15, plancksz15} to a number of \cmmnt{$\sim$15\%} mass measurements on different cluster samples by the South Pole Telescope (SPT) and Planck experiments \citep{geach17,baxter18,raghunathan17b,raghunathan18}. 
Using these techniques, the upcoming CMB experiments, like Simons Observatory \citep{so18} and the proposed CMB-S4 \citep{cmbs4-sb1} experiment are expected to determine cluster masses at 3\% and 1\% level respectively \citep{raghunathan17a}. 

While the CMB lensing signal from large-scale structure is relatively robust to systematics, measurements of the CMB-cluster lensing signal must deal with the thermal Sunyaev-Zel'dovich (SZ) signal from each cluster, which is an order of magnitude greater than lensing signal. 
Quadratic estimators (QE) for the lensing signal are based on the lensing-induced correlations between the large-scale gradient and small-scale CMB anisotropies \citep{hu07}. 
However, the cluster's own SZ emission also creates correlations between these two maps, and can therefore bias the  resulting mass measurements. 

The SZ bias can be side-stepped by either restricting the lensing estimator to CMB polarization data (as the SZ signal is largely unpolarised) (\cite{yasini16,hall14}), or by using a linear combination of temperature maps at different frequencies to remove the SZ signal \citep{baxter15}. 
However, both approaches significantly increase the uncertainties on the lensing reconstruction as, in the first case, the lensing signal is fainter in polarization than temperature, and in the second case, the linear combination increases the final map noise level. 

More recently, \citet{madhavacheril18} and \citet{raghunathan18} demonstrated a modified version of the QE that largely avoids the noise penalty by using the SZ-free map to estimate the large-scale gradient and a single frequency map for the second leg (small-scale lensing map). 
As the SZ signal is only present in one of the two legs, there is no SZ-induced correlation (and thus bias) between the two maps. 
Given the redness of the CMB power spectrum, the large-scale gradient is still measured at high S/N even with the higher noise level of the SZ-free map. 
However, while using a single frequency map reduces the instrumental noise in the small scale lensing map, there remains a S/N penalty due to the extra sample variance from the SZ signal. 
For very massive clusters (or low instrumental noise levels), the SZ variance dominates \citep{raghunathan18} over other sources of uncertainties. 

In this paper, we introduce a refinement on the modified QE that reduces the SZ variance by fitting and removing an SZ template in the second leg. 
We show that the template does not need to be a perfect match to the true cluster signal to substantially reduce the SZ power in the map and improve the final mass estimate. 

The paper is organised as follows.
In \S\ref{sec_simulations}, we describe the simulations we have used to test the proposed method. 
We follow this in \S\ref{sec_methods} by presenting the template fitting refinement to the modified quadratic estimator. 
We report on the performance and robustness of the refined estimator in \S\ref{sec_results}, before laying out forecasts in \S\ref{forecasts} for the improvement in cluster mass constraints from the new estimator for upcoming CMB surveys. 
We conclude in \S\ref{conclusions}.
Throughout, we quote galaxy cluster masses in terms of $M_{200m}$, the mass within a sphere where the density is at least 200 times the mean density of the Universe at that redshift.

\section{Simulations}
\label{sec_simulations}

Simulations of the CMB-cluster lensing signal and a cluster's SZ emission are essential to evaluate the performance of a CMB-cluster lensing estimator. 
The simulations in this work are generated similarly to \citet{raghunathan17a}, and we refer the reader to that work for full details. 
Briefly, for each cluster we create \mbox{$100^\prime \times 100^\prime$} unlensed Gaussian realizations of the \textit{Planck} 2015 \citep{planck15-13} best-fit lensed CMB power spectrum obtained from CAMB\footnote{\url{https://camb.info/}} \citep{lewis00}.
These maps are then lensed by the cluster's convergence profile (see \S\ref{sec_NFW}). 
We add simulated SZ emission (\S\ref{sec_tsz}) and then convolve the map by the assumed instrumental beam. 
Unless otherwise noted, we take the instrumental beam to be Gaussian with a FWHM $=1^\prime.7$ at 90 GHz and $=1^\prime.0$ at 150 GHz. 
Finally, we add Gaussian realizations of white noise at the specified noise level.

\subsection{Lensing convergence signal}
\label{sec_NFW}
We assume the density profile of the galaxy clusters follows the Navarro-Frenk-White (NFW) profile \citep{navarro96}. 
For a  spherical symmetrical density profile like NFW, the resulting lensing convergence, $\kappa(\theta)$, as a function of the angular distance, $\theta$, away from the center is the ratio of surface mass density, $\Sigma(\theta)$, over the critical surface density of the Universe at that redshift, $\Sigma_{crit}(z)$:
\begin{equation}
\kappa(\theta) = \frac{\Sigma(\theta)}{\Sigma_{crit}(z)}.
\end{equation}
We use the closed form expression for NFW lensing convergence presented by \citet{bartelmann96} in this work.

\subsection{thermal Sunyaev-Zel'dovich signal}
\label{sec_tsz}

Unless otherwise noted, we assume the SZ emission is described by a radially-symmetric  Arnaud profile \citep{arnaud10} expected for a cluster of the specified mass and redshift. 
We do test the robustness of this assumption in \S\ref{subsec:simsz} by using two sets of realistic SZ simulations from \citet{sehgal10} and \citet{takahashi17}. 
The SZ power in the \citep{sehgal10} simulations is scaled down by a factor of 1.75 to better match the measured SZ power in \cite{george15}; the \citet{takahashi17} simulations post-date the SZ measurements and are not rescaled. 
Extracting the appropriate SZ signal from these simulations requires selecting matching galaxy clusters, with a tradeoff between the number of matching clusters in the simulations and the fidelity of the match. 
 We randomly select halos from each simulations list of halos which are within 5\% of the desired mass and $\pm 0.2$ of the desired redshift. 
 We then extract the simulation's SZ map, centered at the reported halo location. 
There can be an offset between the reported halo location and peak of the SZ signal; for the \citet{takahashi17} simulations the RMS offset is of order 0.\arcmin{}5.
 Similar offsets have been reported for real clusters \citep{song12b, saro15}. 
 Note that the extracted SZ maps will often include the SZ signal from more than one halo, unlike the lensing maps which do not include the lensing effect of nearby haloes.

\section{Methods}
\label{sec_methods}
In this section, we review the modified QE \citep{madhavacheril18, raghunathan18}, and then present a refinement  of it designed to reduce the SZ variance for massive galaxy clusters or for low noise surveys. 
We also discuss sources of uncertainties present in the CMB-cluster lensing analysis.

\subsection{Modifying the Quadratic Estimator to remove SZ bias}
\label{sec_mQE}

Gravitational lensing by a compact object like a galaxy cluster creates a small-scale dipole pattern in the direction of the large-scale CMB temperature gradient. 
The QE \citep{hu07} is based on extracting the correlation between the dipole and background gradient to estimate the lensing convergence $\hat{\kappa}_{\bL}$. 
Specifically, the lensing convergence can be estimated from a weighted product of filtered versions of a gradient map $G(\hat{n})$ and  small-scale lensing map $L(\hat{n})$: 
\begin{equation}
\hat{\kappa}_{\bL} = -A_{\ell} \int d^{2}\bnhat\ e^{-i\bnhat\cdot\bL}\ {\rm Re} \left\{ \nabla \cdot \left[ G(\bnhat) L^{*}(\bnhat)\right]\right\}.
\label{eq_QE_kappa}
\end{equation} 
Here, $\bnhat$ is the pointing unit vector, $\bL$ is angular multipole, and $A_{l}$ is a normalization factor.

While designed to pull out the lensing-induced correlations between large-scale and small-scale CMB anisotropy, the QE is also sensitive to  correlations due to foreground emission.
Of particular concern is  the cluster's own SZ emission, which is typically an order of magnitude larger than lensing signal. 
As pointed out by \citet{hu07}, the magnitude of the SZ bias can be lowered by decreasing the characteristic scale of the low pass filter on the gradient map, which is set to 2000 in the current work. 
However, a stronger low-pass filter obviously reduces the number of modes used to measure the gradient, and thus decreases the signal-to-noise. 
The modified QE instead eliminates this bias by using an SZ-cleaned map for the large-scale gradient map. 
The large-scale gradient map is chosen because the CMB has much more power on large scales, so the noise penalty from SZ removal has minimal impact. 
Note that while multiple foregrounds can be removed in principle, in practice the focus has been on removing the SZ signal. 
This is simply because the SZ signal introduces the largest bias.  
With the SZ signal present in only one of the two maps, there is no SZ-induced correlation between the two maps and no net bias on the reconstruction of the lensing convergence.

However the SZ emission in the small scale lensing map does add noise to the lensing reconstruction. 
Since under self-similarity the SZ flux \citep{kaiser86}, $y$, is expected to scale with  \mbox{cluster mass $M$} as $y \propto M^{5/3}$ while the lensing signal is linear in mass (in 1-halo regime), the additional SZ variance will generally be more important for high-mass clusters.
The SZ variance will also be more important in low-noise surveys, i.e.~when it is larger than the instrumental noise in the convergence map.

\subsection{Template fitting to reduce the SZ variance}
\label{sec_sz_template_fitting}

An obvious way to eliminate this SZ variance is by using SZ-free maps for both the small-scale and gradient maps. 
Of course this would undo the advantages of the modified QE for the instrumental noise in the small scale map. 
One can also reduce this extra variance by projecting out a model template for the SZ signal, and thereby reducing the total amount of SZ power in the small-scale map. 
Given the impossibility in creating a `perfect' SZ template, the template fitting will not completely eliminate the SZ signal in the small-scale map. 
However, template fitting can significantly reduce the SZ power in the small-scale map, and thus reduce the SZ noise penalty on the lensing mass reconstruction.

To be unbiased, we must fit the template to a CMB-free map or account for the information loss from template projection in the normalization factor $A_l$ of Eqn.~\ref{eq_QE_kappa}. 
The latter tactic (correcting $A_l$) might have advantages if one desired to suppress multiple kinds of cluster emission, e.g., dusty and radio galaxies in addition to the thermal SZ effect. 
Projecting a template from the Compton-$y$ map will not help with these other cluster signals. 
However, we expect these other signals to be small compared to the SZ flux for massive galaxy clusters \citep{raghunathan17a}. 
In this work, we take the first approach and fit the template to a Compton-$y$ map created from a linear combination of 95 and 150\,GHz maps. 
Effectively, we have used the different spectral dependence of the SZ and CMB to eliminate the CMB, and thus any correlation between the removed template and the lensing dipole in the CMB map.

\begin{figure}
\includegraphics[width=\linewidth, keepaspectratio]{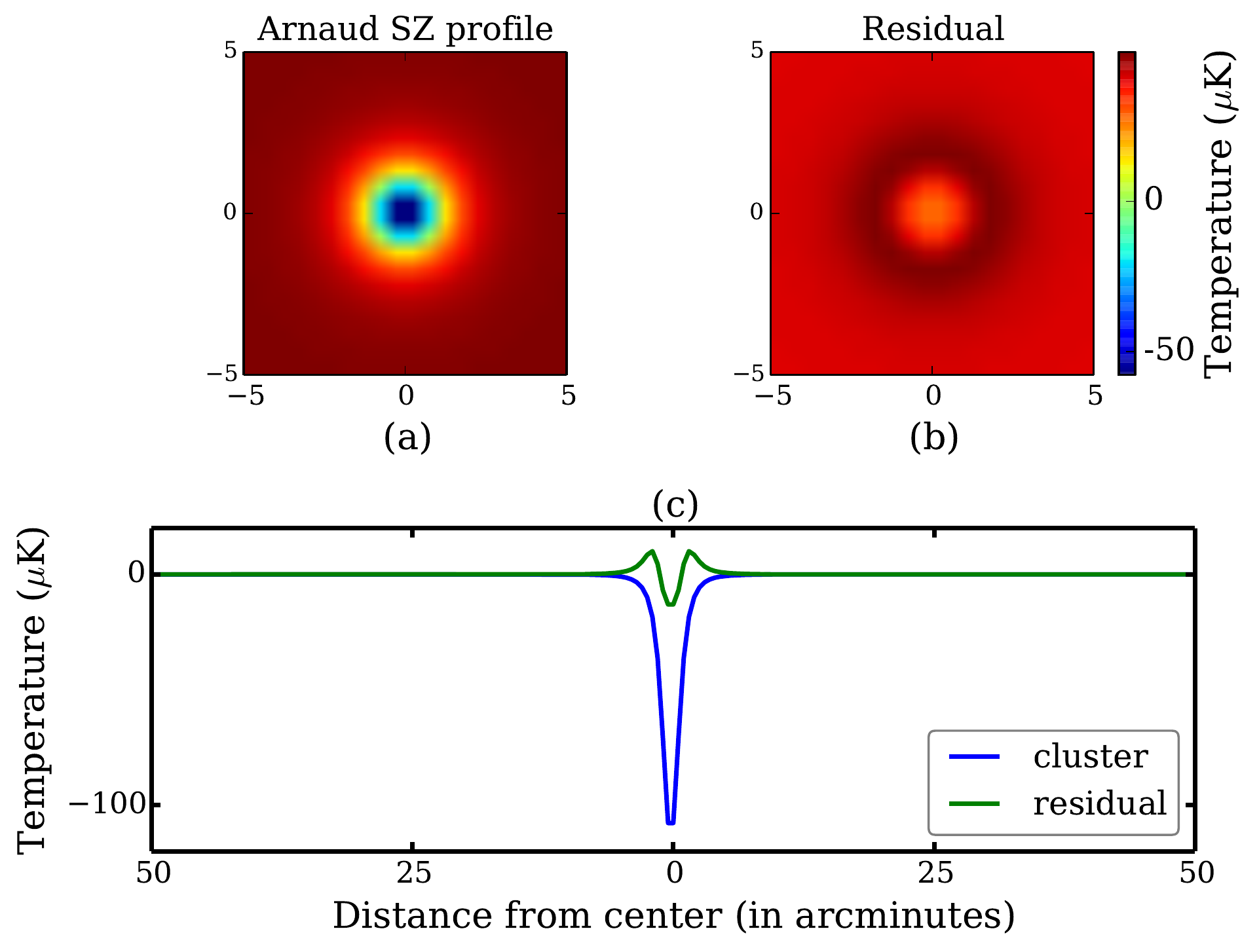}
 \caption{Template fitting significantly reduces SZ power, even with an imperfect match between the template and true SZ signal. 
The top left panel (a) shows the expected Arnaud profile for a galaxy cluster of mass $\mvir = 5 \munits$ at z=0.7 after being smoothed by Gaussian beam with FWHM= 1\arcmin.7.
The top right panel (b) shows the residuals after subtracting the best-fit 2\arcmin.0 FWHM Gaussian (the amplitude is free, but the FWHM is fixed). 
The lower panel (c) shows one-dimensional slices through each panel: the solid, blue line is a slice through the beam-convolved Arnaud profile of (a), and the dashed green line is a slice through the residual map in (b). 
 } 
\label{fig:residual}
\end{figure}

We choose to use a radially symmetric template at the fixed cluster location in this work.  
As the baseline beam size, a Gaussian with FWHM = 1\arcmin{} at 150\, GHz and 1\arcmin.7 at 90\,GHz, is significant compared to the actual size of clusters at $z>0.3$, clusters are approaching the effective point source limit where the specific details of their shape would not matter. 
Thus we choose to use a simple Gaussian template as the baseline template in this work. 
We also compare the results to template fitting with more physically motivated Arnaud profile \citep{arnaud10}, convolved by the experimental Gaussian beam. The residuals of removing a 2\arcmin.0 FWHM Gaussian from an Arnaud profile convolved by a 1\arcmin.7 beam\footnote{The proposed FWHM of SPT-3G 95\,GHz channel is 1.\arcmin 7 and convolving an Arnaud profile with 1.\arcmin 7 beam roughly results in a Gaussian of FWHM 2.0\arcmin{}.} are illustrated in Fig.~\ref{fig:residual}.
The Gaussian template significantly reduces the SZ signal despite the mismatch between the assumed profile and input SZ model. 
While there should be small variations in the typical size of the cluster's SZ emission with mass and redshift, we neglect these variations and fix the size of the templates based on the expected median mass and redshift of the sample. 
However, one could easily adjust the template based on individual cluster redshift for a minimal increase in complexity.

As the SZ effect and CMB have different frequency dependencies, we can use a linear combination of the 95\, GHz and 150\, GHz channels to obtain Compton-$y$ maps.
We then pull out a $10\arcmin \times 10\arcmin$ cutout at the cluster location for fitting the template. 
Since both the SZ emission and lensing signal extraction is concentrated within a few arcminutes of the cluster center, there is little reason to fit over a larger area. 
We allow for two free parameters in the fitting: the overall amplitude of the template, and a constant DC background level. 
While we include the DC term while fitting the template amplitude, we do not subtract DC level. 
Only the template is subtracted from the small-scale map.

With template fitting included, the Fourier transforms of the two maps used by the quadratic estimator can be written down as:
\begin{eqnarray}
G_{\ell} &=& i\ell W^{G}_{\ell} T^{\rm SZ-free}_{\ell}\\
L_{\ell} &=& W^{L}_{\ell} \left(T_{\ell} - T_\ell^{\rm SZ-template}\right)
\end{eqnarray}
where, $G_{\ell}$ is the large-scale gradient map and $L_{\ell}$ is the small-scale map. 
$W^{G}_{l}$ and $W^{L}_{l} $ are the optimal filters to maximize the lensing signal \cite{hu07}. 
The small scale lensing map is $T_{\ell}$ while the constructed SZ-free map is $T^{\rm SZ-free}_{\ell}$. 
Compared to the modified QE \citep{madhavacheril18,raghunathan18}, the new element is the $T_\ell^{\rm SZ-template}$ term representing the SZ template fit.

\subsection{Sources of uncertainty in the CMB-Cluster Lensing Measurement}
\begin{figure}[t]
\includegraphics[width=\linewidth, keepaspectratio]{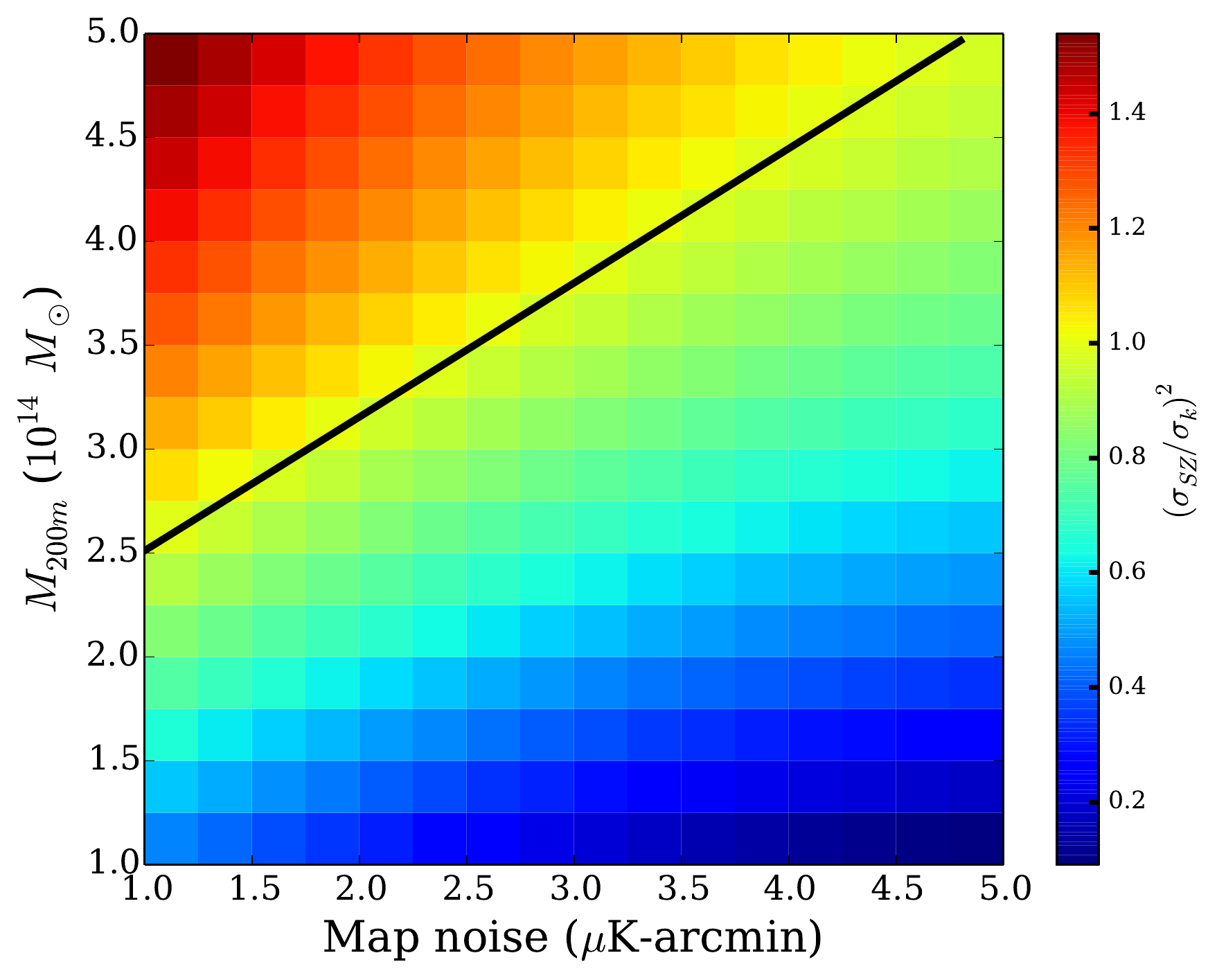}
 \caption{
 Ratio of SZ variance over non-SZ variance as a function of cluster mass and experimental noise level. 
 As expected the ratio increases with mass for a given experimental noise level. 
 The black solid line represents the points where the variance from the SZ and the non-SZ component match contribute equally.
  }
 \label{fig:variance}
\end{figure}

Template fitting is intended to reduce the SZ variance, however it will do nothing for other sources of uncertainty such as instrumental noise, CMB sample variance and foreground emission (if not cleaned). 
Thus it will be useful to look at the relative magnitudes of these two terms, the SZ variance ($\sigma_{SZ}^{2}$) and the non-SZ variance ($\sigma_{\kappa}^{2}$), when interpreting the performance of template fitting in the next section.

The non-SZ variance depends on the survey parameters (i.e.~instrumental noise, and the degree to which foregrounds are cleaned) but is independent of the cluster properties. 
We estimate the non-SZ variance, $\sigma_{\kappa}^{2}$, using simulations of the CMB plus instrumental noise. 
Given the goal, we do not include the cluster's SZ emission.  
We apply the lensing pipeline to the simulated skies to estimate the convergence maps. 
We fit the convergence map to determine the lensing mass, and take the average width of the inferred mass distribution over 1000 simulations to be $\sigma_{\kappa}$.

In contrast, the SZ variance should increase with cluster mass, $M$, roughly as $\sigma_{sz}^2 \propto M^{5/3}$, while being independent of the survey parameters. 
We estimate the SZ variance using the same suite of simulations, however now adding the cluster's SZ emission to the small-scale lensing map. 
We continue using an SZ-free large-scale gradient map.
As before, we estimate the convergence maps, fit for masses and take the average width of the inferred mass distribution to estimate $\sigma_{SZ}^{2}$ + $\sigma_{\kappa}^{2}$.

We present the ratio of the SZ to non-SZ variances, $\sigma_{SZ}^{2}$/$\sigma_{\kappa}^{2}$, as a function of survey noise level and cluster mass in Fig.~\ref{fig:variance}.  
As expected, the ratio increases with mass at any given noise level. 
The black solid curve represents a ratio of unity when the SZ variance equals the non-SZ variance. 
We expect template fitting to significantly improve the mass uncertainties only for clusters above the black curve. 

\section{Results}
\label{sec_results}

\begin{figure}[t]
\includegraphics[width=\linewidth, keepaspectratio]{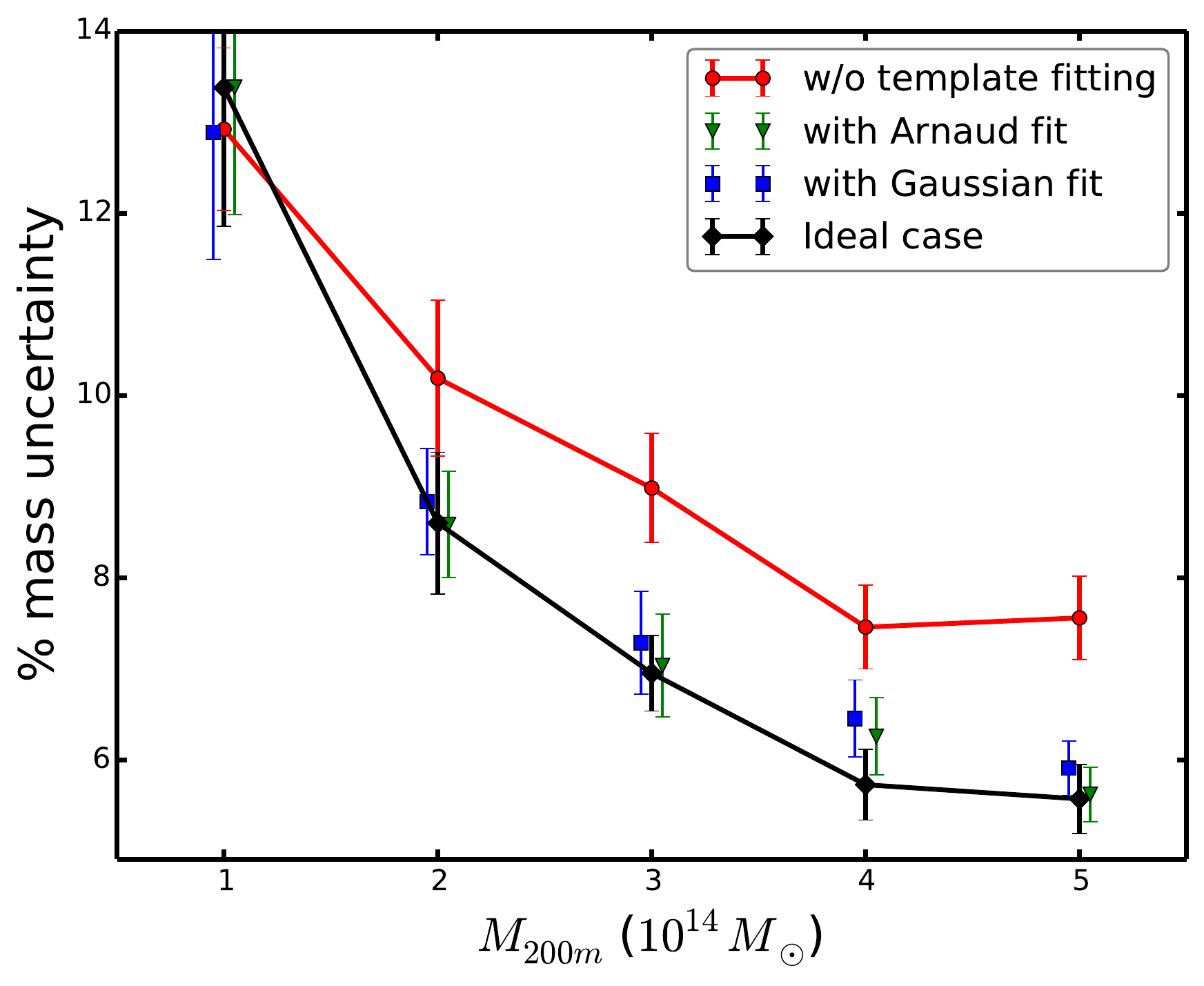}
 \caption{
 Projecting out an SZ template from the second leg of the modified QE improves the estimator's performance for masses above a threshold mass ($2 \times 10^{14}\msolar$ for this specific set of survey parameters).
 Here we show the percentage mass uncertainties from three methods for a stacked sample of 1000 clusters at a experimental noise level of 3\,\ukarcmin{}.
 All the curves use an SZ-free map for the gradient, but make different assumptions about the second, small scale lensing map.
 }
\label{fig:template_fitting}
\end{figure}

The performance of template fitting approach is shown in Fig.~\ref{fig:template_fitting}. 
 We plot the mass uncertainties as a function of mass for the stacked sample of 1000 clusters for an SPT-3G like experiment with FWHM of 1$^\prime$ (1.7$^\prime$) at 150\,GHz (95\,GHz) and a survey noise level of 3\,\ukarcmin{}. In this figure, we compare results for three different small-scale lensing maps (150\, GHz map). 
In all the cases we use a SZ-free gradient map (from a linear combination of 95 and 150\,GHz maps). 
First, the red solid line shows the performance of the original modified QE without template fitting, where we have used 150\,GHz map with Arnaud SZ for the small scale lensing map. 
The second, black solid line shows the results for an idealized case, where we use a 150\,GHz map without SZ for the small scale lensing map. 
While this assumption is unphysical since there will be SZ emission at 150\,GHz, it is useful as a representation of the performance limit for perfect SZ subtraction. 
Finally, the blue squares (green triangles) show the results for the current estimator where we have subtracted a Gaussian (Arnaud) SZ template from the 150\,GHz map.

The relative performance improvement between the idealized (black solid line) and baseline (red solid line) case increases with mass as expected. 
For these survey parameters, the idealized case has 15\% smaller uncertainties than the baseline for clusters of mass $2\times 10^{14}\,\msolar$ and 35\% smaller uncertainties for cluster masses of $5\times 10^{14}\,\msolar$. 
Template fitting recovers nearly all of this gain for high-mass clusters, as shown by the green triangles and blue squares (we have introduced a small offset along mass axis for clarity). 
The green triangle (blue square) shows the performance of projecting out an Arnaud SZ template (Gaussian template) from the small-scale map. 
There is no practical performance difference between the two templates. 
For these assumed survey parameters, both templates are essentially indistinguishable from the idealized perfect SZ removal case at masses above $2\times 10^{14}\,\msolar$. 
Template fitting does not do as well at lower masses. 
This can be understood by considering  the zero-mass limit -- where one tries to remove the SZ template from a map without SZ emission. 
The noisy estimate of the SZ template amplitude will set a non-zero, effective floor to the apparent SZ-like signal in this map. 
In the very low-mass limit, SZ template removal will thus perform more poorly than the original estimator. 
The first signs of this transition can be seen in Fig.~\ref{fig:template_fitting} when comparing the performance at $\mvir = 2\times 10^{14}\,\msolar$ to   $1\times 10^{14}\,\msolar$.


 
\begin{figure}[t]
\includegraphics[width=\linewidth, keepaspectratio]{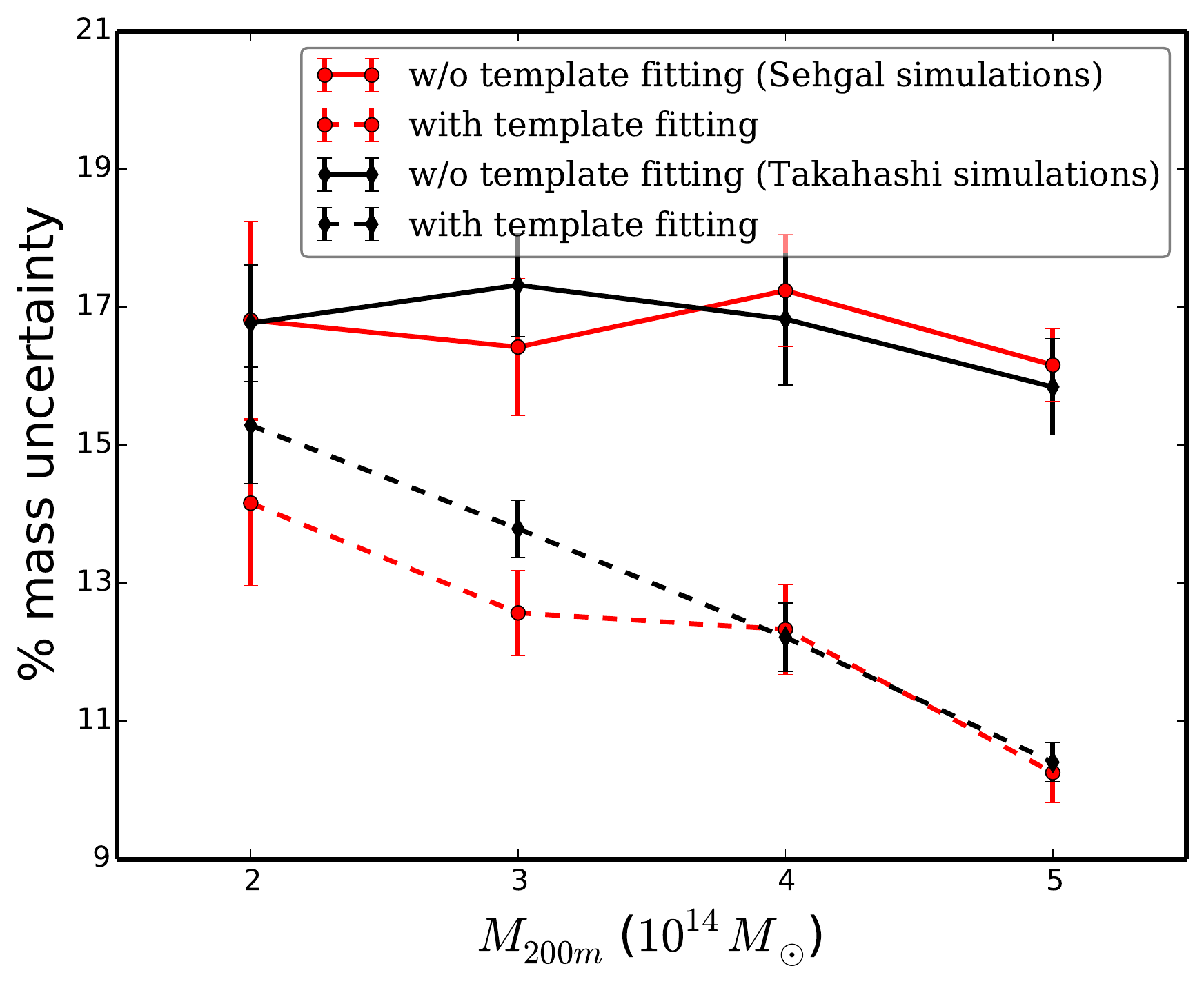}
 \caption{
Our new method is robust to realistic SZ simulations. 
The red and black solid curves show the performance of modified QE for the Sehgal and Takahashi simulations respectively for a stacked sample of 1000 clusters. 
The dashed lines show the improvement in each case when the Gaussian template fitting is used to reduce the SZ variance in the small-scale map. 
The plotted uncertainties are for a stack of 1000 galaxy clusters of the quoted mass at $z=0.7$. 
 }
\label{fig:realistic_sims}
\end{figure}

\subsection{Robustness of template fitting method}
\label{subsec:simsz}

As shown in the last section, the refinement to the modified QE improves the fractional mass uncertainties significantly (above a mass threshold that depends on the survey noise) for the symmetric Arnaud profile. 
However, the question remains as to whether this improvement will continue with the more complex real SZ emission of galaxy clusters along with any other haloes along the line-of-sight.
To test this, we turn to the SZ simulations of \citet{sehgal10, takahashi17}. 
As noted in \S\ref{sec_tsz}, we draw SZ maps from these sims centered at the locations of similarly massed haloes. 
The results for these sims are shown in Fig.~\ref{fig:realistic_sims} for Gaussian template fitting. 
We have tested using both the Gaussian and Arnaud template fitting, and have found no appreciable difference. 
Red lines are for the Sehgal simulations and black lines for the Takahashi simulations. 
The solid lines are without template fitting, while the dashed lines show the improvement from template fitting in each case. 
 It is noteworthy that the mass uncertainties for both sets of simulations are larger at all masses than the case in \S\ref{sec_results} using the NFW mass profile and Arnaud SZ profile. 
 The increased uncertainties are due to the overall level of SZ power as well as additional scatter from other haloes nearby or along the line of sight. 
 Regardless, the fractional improvement is significant -- a factor of 1.6 reduction of mass uncertainties for cluster masses of $5\times 10^{14}\,\msolar$. 
 This is nearly identical to (even slightly better than) the factor of 1.5 reduction seen for these clusters in \S\ref{sec_results}.
 We conclude that template fitting works well even for more complicated and realistic SZ profiles.


\section{Forecasts}
\label{forecasts}

We now consider the impact of this method on the performance of upcoming CMB surveys, SPT-3G \citep{bender18}, Simons Observatory \citep{so18} and CMB-S4 \citep{cmbs4-sb1}. 
For the latter two experiments, we only consider the large-aperture telescopes which are expected to have experimental beams of $\theta_{\rm FWHM} = 1.^{\prime}4$ at 150 GHz and to cover 40\% of the sky (see also Table \ref{table_forecast_setup}). 
For SPT-3G, we assume a survey area of 1500\,\sqdeg{} and an experimental beam of $\theta_{\rm FWHM} = 1.^{\prime}2$ at 150 GHz. 
We create a single cluster catalog realization for each experiment based on the noise levels at 150\,GHz. 
Note that for simplicity, we make no attempt to use frequency information to remove other temperature foreground signals such as the the CIB, nor to improve the noise level on the Compton-$y$ map, even though such techniques might allow more clusters to be detected.

The simulated 150\,GHz maps used for cluster finding include the primary CMB anisotropy, Gaussian foregrounds uncorrelated with the clusters \citep{george15}, cluster SZ emission modeled using Arnaud profile, and the respective white noise levels as given in Table \ref{table_forecast_setup}.
Next, we use the publicly available\footnote{\url{http://hmf.icrar.org/}} Halo Mass Function calculator \citep{murray13} to obtain the halo counts $d^{2}N/(dz\ d{\rm log}M$) per unit redshift ($\Delta z = 0.1$) and mass ($\Delta {\rm log} M = 14.05$) bin. 
Using the results from the above simulations, we get the number of clusters that will be detected above $\snr\ge 5$ by the three experiments. 
This detection $\snr$ threshold is extremely conservative as well (the sample purity should be close to 1). 
We estimate that SPT-3G, Simons Observatory and CMB-S4 will detect approximately 2,400, 27,000 and 75,000 clusters respectively under these simplifying assumptions.

The three simulated cluster sample and assumed experimental parameters are passed through the cluster-lensing pipeline to estimate the mass uncertainty from CMB-cluster lensing on each stacked sample. 
 For these forecasts we draw SZ profiles from the \cite{takahashi17}  simulations for haloes of similar mass and redshift (see \S\ref{sec_tsz}).
The results are given in Table \ref{table_forecast_setup}. 
We find that template fitting reduces the mass uncertainty for SPT-3G, Simons Observatory and CMB-S4 by factors of 1.39, 1.33 and 1.46 respectively.

\begin{table}
\caption{Template fitting will improve the CMB-cluster lensing mass calibration for upcoming experiments. Here we list forecasts for SPT-3G, Simons Observatory (SO), and CMB-S4}
\footnotesize{
\centering
\begin{tabular}{| C{1.34cm} | C{1.45cm}|  C{0.9cm} | C{1.6cm} | C{1.6cm}  |}
\hline
 \multirow{3}{*}{Experiment} &  \multirow{2}{*}{Noise $\Delta_{T}$ } & \multirow{3}{*}{N$_{\rm clus}$} & \multicolumn{2}{c|}{$\Delta M/M\ [\%] $(for 1000 clusters)}\\
\cline{4-5}
 & at 150GHz [\ukarcmin] & & No template fitting & With template fitting\\\hline
SPT-3G & 2.2 & 2,400 &11.1& 8.0 \\\hline
SO & 6.0 & 27,000 &2.8 &2.1 \\\hline
CMB-S4 & 1.7 & 75,000 &1.9 & 1.3\\\hline

\end{tabular}
}
\label{table_forecast_setup}
\end{table}

\section{Conclusion}
\label{conclusions}

We have presented an improved version of the QE to estimate galaxy cluster masses through CMB lensing. 
By projecting out an SZ template from the high-pass filtered leg of the QE, the algorithm substantially reduces the variance due to a cluster's own SZ emission. 
The SZ variance can be significant fraction of the total variance for high-mass clusters or low-noise surveys. 
Note that in the opposite limit of low masses or high noise (i.e. when the SZ variance is negligible), this template fitting may slightly reduce the overall S/N. 

While the performance of the template fitting does depend on the fidelity with which the template accurately represents the true SZ signal, these variations do not seem to be a significant issue. 
In part this is because upcoming CMB experiments will not have sufficient angular resolution to resolve most of the structure in the SZ emission; the instrumental beam at 150\,GHz for a 6\,m telescope is slightly larger than a typical high-redshift cluster. 
For mock SZ signals modeled using Arnaud profile, we find that the template removal approach eliminates all of the variance due to the SZ signal. 
The improvement is not 100\% for realistic SZ signals \citep{sehgal10, takahashi17} because of the following reasons: (1) nearby haloes add some SZ signal, (2) the radially symmetric template can not remove the non-symmetric portion of the cluster's SZ signal, and (3) there can be an offset between the centers of the template and true SZ signal.

The template fitting approach may also be useful for estimating the CMB lensing power spectrum. 
\citet{madhavacheril18} has already demonstrated that the modified QE is useful to reconstruct the lensing convergence of the large-scale structures. 
In the context of the lensing auto-spectrum, the additional SZ variance from galaxy clusters will increase the magnitude of the noise bias correction. 
This has historically been avoided by masking the positions of massive galaxy clusters in the map, but the masking approach becomes impractical as the number of bright clusters per square degree increases. 
Given the numbers of clusters that will be found by the Simons Observatory and CMB-S4, it will be worthwhile to explore alternatives to masking. 
One possibility would be to inpaint the background CMB at the cluster locations \citep{benoitlevy13}.
One might instead use the template fitting approach in this work would reduce the magnitude of the noise bias due to the SZ variance, although  one would need to consider carefully how this approach affected the final uncertainty on the bias correction. 
We leave a detailed investigation of this to a future work.

CMB-cluster lensing will be a key tool for galaxy cluster cosmology at high redshifts, while also providing a cross-check of optical weak lensing mass estimates at lower redshifts. 
For future surveys like CMB-S4 and Simons Observatory and for clusters above masses of $4\times 10^{14}\,\msol$, we find that template fitting yields a factor of 1.4 reduction in the mass uncertainties. 
Better cluster mass measurements will help us in the quest to understand dark energy and the accelerating expansion of the Universe.

\subsection*{Acknowledgements}
We thank Eric Baxter, Federico Bianchini, Thomas Crawford, Nikhel Gupta, Daisuke Nagai, and W. L. Kimmy Wu for useful discussions regarding this project.

SP acknowledges support from MIPP, Laby Travel Bursary, and Melbourne International Engagement Award. 
SR acknowledges the support from NSF grants AST-1716965 and CSSI-1835865.
The work at Melbourne is supported by the Australian Research Council's Discovery Projects scheme (DP150103208).
We thank the high performance computation center at the University of Melbourne for providing access to the cluster \texttt{spartan.unimelb.edu.au}.
We acknowledge the use of \texttt{CAMB} \citep{lewis00} and \texttt{HEALPIX} \citep{gorski05} routines. 
This research used resources of the National Energy Research Scientific Computing Center (NERSC), a DOE Office of Science User Facility supported by the Office of Science of the U.S. Department of Energy under Contract No. DE-AC02-05CH11231. 

\bibliographystyle{aasjournal}
\bibliography{spt}

\end{document}